\documentstyle[12pt]{article}
\newcommand{\beq}{\begin{equation}}
\newcommand{\eeq}{\end{equation}}
\newcommand{\beqq}{\begin{eqnarray}}
\newcommand{\eeqq}{\end{eqnarray}}
\begin{document}
\begin{center}
{\bf Exact ground and excited states of frustrated
antiferromagnets on the $CaV_4O_9$ lattice}
\vskip 1cm
Indrani Bose and Asimkumar Ghosh
\\Department of Physics,\\Bose Institute,\\
93/1, Acharya Prafulla Chandra Road,\\Calcutta-700 009, India.
\end{center}
\begin{abstract}
The experimental observation of a spin-gap in the antiferromagnetic 
spin-$\frac{1}{2}$ compound $CaV_4O_9$ has been attributed to the 
formation of a plaquette resonating-valence-bond (PRVB) state in 
the underlying 1/5-depleted square lattice. We construct a 
spin-$\frac{1}{2}$ model on the 1/5-depleted lattice for which the PRVB 
state can be shown to be the exact ground state in a particular 
parameter regime. In a subspace of this parameter regime, the first 
excited state can be calculated exactly. In a different 
parameter regime, the dimer state is shown to be the exact ground state.
\end{abstract}

\vskip 1cm

Recent discovery of a spin-gap in $CaV_4O_9$\cite{Taniguchi} has
given rise to a number of studies\cite{Ueda,Katoh,Sano,Albretch,
Troyer,Starykh,Gelfand} to understand the
origin of the gap. The magnetic system can be described by
the antiferromagnetic (AFM) Heisenberg Hamiltonian for spins of
$V^{4+}$ ions $(S=\frac{1}{2})$ on a 1/5-depleted square
lattice. The depleted lattice is obtained from the original
square lattice by removing $\frac{1}{5}$ of the lattice sites in
a regular manner. The depleted lattice is a bipartite lattice
which consists of 4-spin plaquettes connected by dimer bonds
designated as A and B respectively in Fig. 1.

The minimal model to describe the system is the Heisenberg
Hamiltonian 
\begin{eqnarray*}
H\,=\,\sum_{n.\,n.}\,J_{nn}\,\vec{S}_i\cdot\vec{S}_j
\end{eqnarray*}
for nearest-neighbour (n.n.) spin-spin interactions. The exchange integral 
$J_{nn}$ equals to $J_0\,(J_1)$ for plaquette (dimer) bonds. Two limiting 
cases that can be considered are: (i) $J_0\,\gg\,J_1$ and 
(ii) $J_1\,\gg\,J_0$. In the first case, the ground state is the 
plaquette-resonating-valence-bond (PRVB) state in which the spin 
configuration in each plaquette is of the RVB type. In the other limit, 
the ground state consists of singlets (dimers) along the dimer bonds. 
In both the cases the ground state is a singlet and spin-disordered. 
Also, the spin-gap has a non-zero value. As pointed out by Ueda et al
\cite{Ueda}, the wave function of the dimer phase is orthogonal to that 
of the PRVB phase. The reason is that the two wave functions have 
different symmetry properties under reflection with respect to the dimer 
bonds: odd for the dimer state and even for the PRVB state. Thus the two 
states cannot be continuously connected by changing the parameters $J_0$ 
and $J_1$. An intermediate phase is expected to occur between the dimer 
and PRVB phases. This phase has been shown\cite{Troyer} to have long range 
N\'{e}el-type order and includes the isotropic coupling case $J_0\,=\,J_1$. 
The spin-gap defined to be the difference in the energies of the 
lowest triplet excited state and the ground state, vanishes in the N\'{e}el ordered 
phase. Thus isotropic or near-isotropic n.n. coupling cannot explain the 
origin of the spin-gap.

Spin-gap has been suggested to occur when $J_0\,\neq\,J_1$ or due to the 
inclusion of frustrated next-nearest-neighbour (n.n.n.) couplings\cite{Ueda}. 
It has been further suggested\cite{Starykh} that the spin-Peierls mechanism 
in the $CaV_4O_9$ lattice strengthens the plaquette bonds with respect to 
the inter-plaquette bonds. Taking these suggestions into account, 
we construct an AFM Heisenberg Hamiltonian on the 1/5-depleted lattice 
for which the exact ground states in different parameter regimes can be 
determined. These ground states include both the RVB and dimer states. 
In addition, a number of excited states are determined 
exactly. 

The Hamiltonian includes the following types of interactions: 
n.n. plaquette, plaquette diagonal, dimer bonds connecting plaquettes, 
next-nearest-neighbour (n.n.n.), 
 diagonal bonds connecting plaquettes 
and Knight's-move-distance-away (KM) 
interactions of strength $J_1,\,J_2,\,J_3,\,J_4,\,J_5$ and $J_6$ respectively. 
The pairwise interactions are illustrated in Fig. 2. The n.n.n. bonds 
connecting the mid-points of the big squares have strength zero. Similarly, 
the KM bonds which lie outside plaquettes have strength zero.  
The unequal interaction strengths are possible due to the 
spin-Peierls mechanism. The exchange interaction Hamiltonian is given by 
\begin{equation}
\label{eqn. 1}
H_{AF}\,=\,\sum_{<ij>}\,J_{ij}\,\vec{S}_i\cdot\vec{S}_j 
\end{equation}
For the ratio of interaction strengths
\beq
\label{eqn. 2}
J_1\,:\,J_2\,:\,J_3\,:\,J_4\,:\,J_5\,:J_6\,=\,1\,:\,\beta\,:\,\alpha\,:\,
\frac{\alpha}{2}\,:\,\frac{\alpha}{2}\,:\,\frac{\alpha}{2}
\eeq
the PRVB state can be shown to be the exact ground state, when 
$\beta\,\leq\,1$ and $\alpha\,\leq\,.5$.

The PRVB state has a 
RVB spin configuration in each plaquette. The RVB configuration is a 
linear superposition of two valence bond (VB) configurations (Fig. 3). 
A valence bond is a spin singlet and is represented by a solid line in the 
Figure. The arrow on the solid line specifies the ordering of spins 
in the singlet, e.g.,
\setlength{\unitlength}{0.240900pt}
\ifx\plotpoint\undefined\newsavebox{\plotpoint}\fi
\begin{picture}(300,100)(700,560)
\put(750,534){\makebox(0,0)[l]{i\quad\quad\,\,\,j}}
\put(812,585){\makebox(0,0)[l]{$>$}}
\put(750,585){\usebox{\plotpoint}}
\put(750,585){\raisebox{-.8pt}{\makebox(0,0){$\bigcirc$}}}
\put(889,585){\raisebox{-.8pt}{\makebox(0,0){$\bigcirc$}}}
\put(750.0,585.0){\rule[-0.500pt]{33.485pt}{1.000pt}}
\end{picture}
denotes the spin configuration 
$\frac{1}{\sqrt{2}}\left(\alpha (i)\,\beta (j)\,-\,\beta (i)\,\alpha (j)\right)$;
$\alpha$ represents an up-spin and $\beta$ a down-spin, i and j are 
the lattice sites. The PRVB state is further an exact eigenstate of $H_{AF}$ 
as can be easily verified by using any one of the following spin identities
\begin{eqnarray}
\label{eqn. 3}
\vec{S}_n\cdot (\,\vec{S}_l\,+\vec{S}_m\,)\,[lm]\,\equiv\,0 \;\;\;\;\;\;\;\;\;\;\;\;(3a)\nonumber\\
\vec{S}_n\cdot(\,\vec{S}_l\,+\,\vec{S}_m\,+\,\vec{S}_p\,+\,\vec{S}_q\,)[lmpq]\,\equiv\,0\;\;\;\;\;\;\;(3b)\nonumber
\end{eqnarray}
where [lm] denotes the spin singlet associated with a bond lm and
[lmpq] is the singlet ground state (Fig. 3) of a plaquette lmpq of four spins. 
Each plaquette is part of two octahedra 
as shown in Fig. 4. The sites i, j, k, o, n, p form one octahedron and 
the sites l, j, k, o, n, m form the other octahedron (the lines of the 
octahedra are not joined for clarity). When the plaquette is in a RVB state, 
as shown in Fig. 3, the non-plaquette interactions of the octahedra, 
corresponding to the bonds ij, ik, in, io, po, pn, pk, pj and 
lk, lo, ln, lj, mn, mj, mk, mo give zero contribution to the energy because of 
the spin identities in (3). 
 The PRVB state is thus the exact 
eigenstate of $H_{AF}$. The exact energy per plaquette is 
$-2\,J_1\,+\,\frac{J_2}{2}$. To prove that the exact eigenstate is also 
the exact ground state, we make use of the method of `divide and conquer'
\cite{Anderson}. Let $E_1$ and $E_g$ be the exact eigenstate and ground state 
energies. Thus $E_g\,\leq\,E_1$. Let $\Psi_g$ be the exact ground state 
wave function. The Hamiltonian $H_{AF}$ is divided into sub-Hamiltonians 
$H_i$'s describing the octahedra of interactions, $H_{AF}\,=\,\sum_i\,H_i$. 
Let $E_{ig}$ be the ground state energy of $H_i$.

For an octahedron, let $J_1^\prime,\,J_2^\prime,\,J_3^\prime,\,
J_4^\prime,\,J_5^\prime,$ and $J_6^\prime$ be the strengths of the various 
interactions. The primed symbols denote the same types of interactions 
as the unprimed symbols do. Let $J_2^\prime\,=\,\beta\,J_1^\prime$ and  
$J_3^\prime\,=\,J_4^\prime\,=\,J_5^\prime\,=\,J_6^\prime\,
=\,\alpha\,J_1^\prime$. When $\beta\,\leq\,1$ and $\alpha\,\leq\,\frac{1}{2}$, 
the exact ground state of the octahedron consists of the RVB state of Fig. 3 
in the plaquette and free vertex spins which can be either up or down. 
The exact ground state energy 
$E_{ig}\,=\,-2\,J_1^\prime\,+\,\frac{J_2^\prime}{2}$. When the 
sub-Hamiltonians  $H_i$'s are added, all the plaquette bonds and the dimer 
bonds connecting the plaquettes are counted twice. So, 
$J_1\,=\,2\,J_1^\prime,\,J_2\,=\,2\,J_2^\prime,\,J_3\,=\,2\,J_3^\prime,\,
J_4\,=\,J_4^\prime,\,J_5\,=\,J_5^\prime$ and $J_6\,=\,J_6^\prime$. 
Using the variational theorem, 
\begin{eqnarray*}
E_g\,&=&\,\langle\,\Psi_g\,|\,H_{AF}\,|\,\Psi_g\rangle 
     =\,\sum_i\,\langle\,\Psi_g\,|\,H_i\,|\,\Psi_g\,\rangle\, \\
    &\geq&\,\sum_i\,E_{ig}\,\geq\,N_p\,(\,-4\,J_1^\prime\,+\,J_2^\prime\,)\,
     =\,N_p\,(\,-2\,J_1\,+\,\frac{J_2}{2}\,)
\end{eqnarray*}
Where $N_p$ is the number of the plaquettes in the lattice. The PRVB state 
which is the exact eigenstate of the $H_{AF}$ has the energy 
$E_1\,=\,N_p\,(\,-\,2\,J_1\,+\,\frac{J_2}{2}\,)$. Thus in 
$\sum_i\,E_{ig}\,\leq\,E_g\,\leq\,E_1$ the lower bound $\sum_i\,E_{ig}$ has
the same value as the upper bound $E_1$. In other words, the exact PRVB 
eigenstate of $H_{AF}$ is the exact ground state with the ratio of 
interaction strengths $J_1\,:\,J_2\,:\,J_3\,:\,J_4\,:\,J_5\,:\,J_6\,=\,
2\,J_1^\prime\,:\,2\,\beta\,J_1^\prime\,:\,2\,\alpha\,J_1^\prime\,
:\,\alpha\,J_1^\prime\,:\,\alpha\,J_1^\prime\,:\,\alpha\,J_1^\prime$ 
which reduces to the ratio given in (2).

A RVB state similar to that shown in Fig. 3 is obtained by replacing 
the `+' sign by a `$-$' sign. We designate this new RVB state as 
$\Psi_{RVB1}$ with the original RVB state labeled as $\Psi_{RVB}$. 
We construct a new state PRVB1 (total spin $S\,=\,0$) for the 
1/5-depleted lattice in which all the plaquettes, except one, have 
spin configuration corresponding to $\Psi_{RVB}$. In one plaquette, 
the spin configuration corresponds to $\Psi_{RVB1}$. For the ratio of 
interaction strengths given in (2) and when
\beqq
\label{eqn. 4}
\beta\,\geq\,\alpha\,+\,0.5\quad\quad\quad\quad\quad\quad\quad\quad\quad(4)\nonumber
\eeqq
the PRVB1 state can be shown to be exact first excited state of $H_{AF}$.

As in the case of the ground state, the proof can be obtained by means 
of the spin identity in (3) and using the method of `divide and conquer'. 
Exact diagonalisation of an octahedron corresponding to a sub-Hamiltonian 
$H_i$, shows that $\Psi_{RVB}$ is the ground state and $\Psi_{RVB1}$ 
the first excited state when $\beta\,\leq\,1,\,\alpha\,\leq\,\frac{1}{2}$ and 
the restriction in (4) is satisfied. The rest of the proof 
is similar to that for the ground state. The energy of the first 
excited state of $H_{AF}$ is $E_f\,=\,-\,\frac{3\,J_2}{2}\,+\,(N_p\,-\,1)
(-\,2\,J_1\,+\,\frac{J_2}{2})$. The energy of the ground state is 
$E_g\,=\,N_p\,(-\,2\,J_1\,+\,\frac{J_2}{2})$. Then the energy difference D is 
\beqq
\label{eqn. 5}
D\,=\,E_f\,-\,E_g\,=\,2\,J_1\,-\,2\,J_2\quad\quad\quad\quad\quad(5)\nonumber
\eeqq
Evidence of a $S\,=\,0$ first excited state has been obtained earlier in 
the numerical calculations of\cite{Mila}.

The PRVB1 state is $N_p$-fold degenerate as the plaquette, in which the 
spin configuration is $\Psi_{RVB1}$, can be any one of the $N_p$ plaquettes. 
Other excited states can be constructed in which r ($r\,=\,2,\,3,\,4,....$ 
etc.) plaquettes have the spin configuration $\Psi_{RVB1}$. In all the other 
($N_p\,-\,r$) plaquettes the spin configuration is $\Psi_{RVB}$.

We now consider the parameter regime in which the dimer state is the exact 
ground state. The ground state consists of singlets along the dimer bonds. 
It can be shown that the dimer state is the exact ground state when the ratio 
of interaction strengths is 
\beqq
\label{eqn. 6}
J_1\,:\,J_2\,:\,J_3\,:\,J_4\,:\,J_5\,:\,J_6\,=\,\frac{\gamma}{3}\,
:\,\frac{\gamma}{3}\,:\,1\,:\,\frac{\gamma}{6}\,:\,\frac{\gamma}{6}\,:\,
\frac{\gamma}{6}\quad\quad(6)\nonumber
\eeqq
where $\gamma\,\leq\,1$. The method of proof is the same as in the case 
when the PRVB state is the exact ground state. The dimer state can be 
shown to be an exact eigenstate by making use of the spin identity 
(3).

In the dimer ground state, the energy associated with each dimer bond is 
$-\,\frac{3\,J_3}{4}$. While using the method of `divide and conquer', 
the Hamiltonian $H_{AF}$ is sub-divided into Hamiltonian $H_i$'s, each $H_i$ 
describing a triangle of spins. The triangle consists of a dimer bond 
(strength $J_3^\prime$) and two other bonds (strength $\gamma\,J_3^\prime$). 
For $\gamma\,\leq\,1$, the ground state of the triangle is a singlet along 
the dimer bond. Each dimer bond connects two plaquettes and each plaquette 
contributes three sub-Hamiltonians which include the dimer bond in question. 
For example, in Fig. 4, the plaquette jkon contributes three sub-Hamiltonians 
corresponding to the triangle jmn, kmn and omn, all of which include the 
dimer bond mn. When adding all the $H_i$'s, each dimer bond is counted 
six times, the plaquette bonds are counted twice and the other bonds are 
counted once. This leads to the ratio of interaction strengths given in 
(6).

We now discuss the relevance of our model for understanding the 
properties of the experimental system $CaV_4O_9$. Earlier theoretical 
studies \cite{Ueda,Katoh,Sano,Albretch,Troyer,Starykh} 
have explained the experimentally-observed spin gap  
by assuming $CaV_4O_9$ to be in a PRVB phase. Our model AFM 
Hamiltonian shows that in a particular parameter regime the PRVB 
state is the exact ground state on the 1/5-depleted square lattice. 
Our model includes only two more further-neighour interactions 
(n.n.n. and KM) than those considered in the earlier papers. We 
have assigned different interaction strengths to the intra-plaquette 
and inter-plaquette bonds. This is not unrealistic in view of the 
possibility that the spin-Peierls mechanism is at work in $CaV_4O_9$ 
\cite{Starykh,Read}. The ratio of interaction strengths for which 
the exact ground and first excited states can be determined 
corresponds to a region, rather than a single point, in parameter space. 
More experiments are, however, needed to settle the question of how many 
exchange parameters are needed to give a realistic description of $CaV_4O_9$. 

As already mentioned, Albrecht et al\cite{Mila} find the existence of singlet 
states in the spin gap, from the results of exact diagonalisation of finite 
systems. In the parameter regime given in Eq.(4), the PRVB1 state which is a 
singlet (S=0) has energy lower than that of the first triplet excited state. 
This state thus constitutes an in-gap state. Experimental signature 
of this state can be obtained from Raman scattering experiments. The fact 
that the singlet excitation is the lowest excited state is due to the 
presence of further-neighbour interactions. 
In this case, there should be an asymmetry 
\beqq
\Delta (k_x,\,k_y)\,\neq \,\Delta (k_y,\,k_x) \nonumber
\eeqq
in the spin-gap at the wave vector $\vec{k}\,=\,(k_x,\,k_y)$. This 
asymmetry can be detected in neutron scattering experiments. There is so 
far no theory which can give a good fit to the experimental data of the 
susceptibility over the whole temperature range of 20 K to 700 K. 
Two possible reasons that have been suggested are: (i) inclusion of more 
terms in the Hamiltonian besides n.n. and diagonal and (ii) the dependence 
of the exchange constants on temperature due to spin-phonon coupling. 
Our model Hamiltonian includes more interaction terms than in earlier models. 
A theoretical calculation\cite{Asim} on the temperature dependence of 
susceptibility is in progress and the results will be published elswhere.

In summary, we have constructed a $S\,=\,\frac{1}{2}$ AFM Heisenberg model 
on the 1/5-depleted lattice for which the PRVB and dimer states are the 
exact ground states in different parameter regimes. Both these states are 
singlets and spin-disordered. In the parameter regime in which the PRVB 
state is the exact ground state, an added restriction on the parameter 
values, makes it possible to obtain the exact first excited state.
 The relevance of the model to 
understand the physical properties of $CaV_4O_9$ is not well-established. 
The model, however, can serve as a toy model to study quantum phase 
transition from the PRVB to the dimer phase and also can be a starting 
point for studies of more realistic model Hamiltonians.

\newpage
{\Large Figure Captions}

\begin{description}
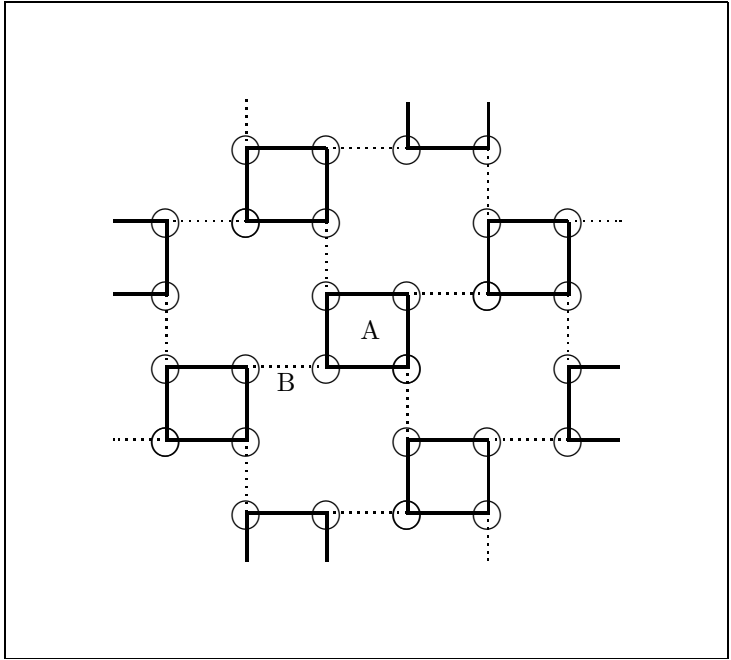
\item [Fig. 1]The 1/5-depleted lattice of $CaV_4O_9$. A and B represent
four-spin plaquettes and dimer bonds connecting plaquettes, respectively.
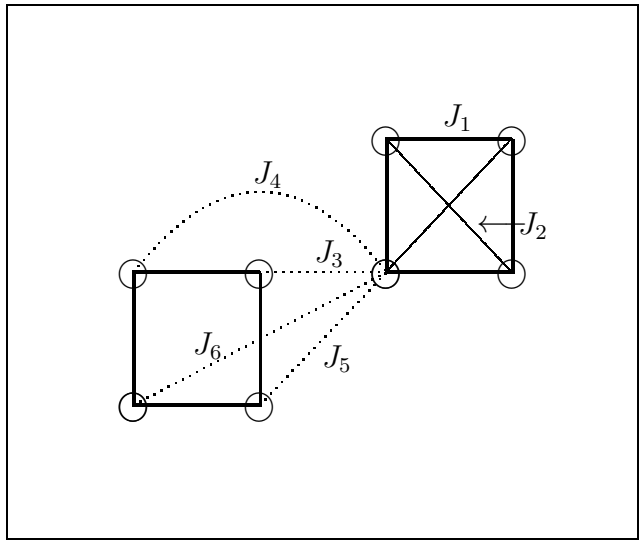
\item [Fig. 2] The n.n. plaquette, plaquette diagonal, dimer bonds connecting
plaquettes, next-nearest-neighbour (n.n.n.), diagonal bonds connecting
plaquettes and Knight's-move-distance-away (KM) interactions of strengths
$J_1,\,J_2,\,J_3,\,J_4,\,J_5$ and $J_6$ respectively. For clarity, the
non-nearest-neighbour bonds are drawn singly.
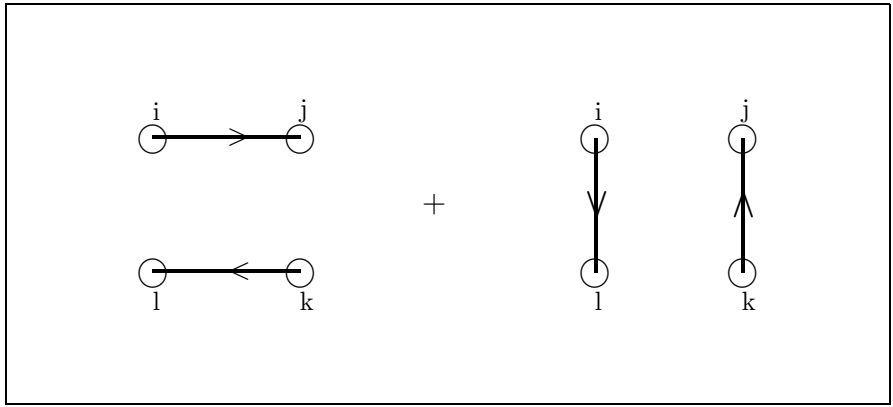
\item [Fig. 3] The RVB spin configuration in each four-spin plaquette in the
PRVB state. The solid line denotes a singlet. The arrow sign indicates the
ordering of spins in a singlet; i, j, k and l are the lattice sites.
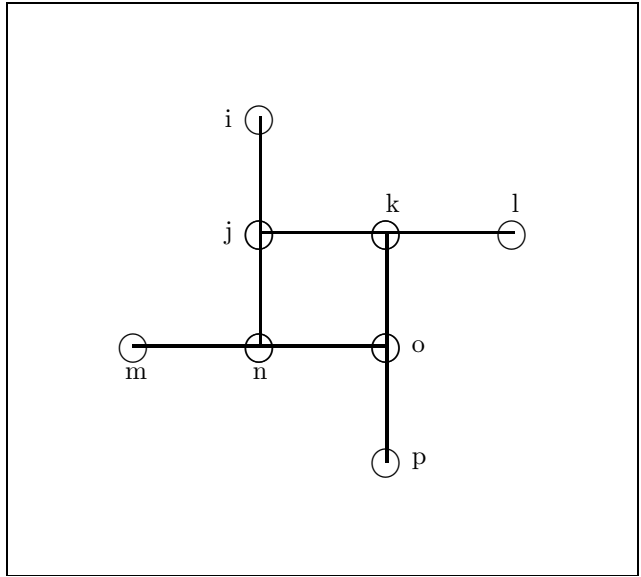
\item [Fig. 4]The plaquette okjn as part of two octahedra. The vertices i, l, p
and m are connected to all the sites of the plaquette. Many of these
connections are not shown in the Figure for clarity.
\end{description}

\newpage
\setlength{\unitlength}{0.240900pt}
\ifx\plotpoint\undefined\newsavebox{\plotpoint}\fi
\sbox{\plotpoint}{\rule[-0.500pt]{1.000pt}{1.000pt}}%
\begin{picture}(900,1300)(90,0)
\font\gnuplot=cmr10 at 10pt
\gnuplot
\sbox{\plotpoint}{\rule[-0.500pt]{1.000pt}{1.000pt}}%
\put(860,698){\makebox(0,0)[l]{A}}
\put(729,617){\makebox(0,0)[l]{B}}
\put(794,125){\makebox(0,0)[l]{{\large Fig. 1}}}
\put(680,411){\usebox{\plotpoint}}
\put(680.0,411.0){\rule[-0.500pt]{30.353pt}{1.000pt}}
\put(554,755){\usebox{\plotpoint}}
\put(554.0,755.0){\rule[-0.500pt]{1.000pt}{27.463pt}}
\put(933,984){\usebox{\plotpoint}}
\put(933.0,984.0){\rule[-0.500pt]{30.353pt}{1.000pt}}
\put(1185,526){\usebox{\plotpoint}}
\put(1185.0,526.0){\rule[-0.500pt]{1.000pt}{27.463pt}}
\put(933,411){\usebox{\plotpoint}}
\put(933.0,411.0){\rule[-0.500pt]{30.353pt}{1.000pt}}
\put(1059.0,411.0){\rule[-0.500pt]{1.000pt}{27.703pt}}
\put(933.0,526.0){\rule[-0.500pt]{30.353pt}{1.000pt}}
\put(933.0,411.0){\rule[-0.500pt]{1.000pt}{27.703pt}}
\put(933,640){\usebox{\plotpoint}}
\put(933.0,640.0){\rule[-0.500pt]{1.000pt}{27.703pt}}
\put(806.0,755.0){\rule[-0.500pt]{30.594pt}{1.000pt}}
\put(806.0,640.0){\rule[-0.500pt]{1.000pt}{27.703pt}}
\put(806.0,640.0){\rule[-0.500pt]{30.594pt}{1.000pt}}
\put(554,526){\usebox{\plotpoint}}
\put(554.0,526.0){\rule[-0.500pt]{30.353pt}{1.000pt}}
\put(680.0,526.0){\rule[-0.500pt]{1.000pt}{27.463pt}}
\put(554.0,640.0){\rule[-0.500pt]{30.353pt}{1.000pt}}
\put(554.0,526.0){\rule[-0.500pt]{1.000pt}{27.463pt}}
\put(680,869){\usebox{\plotpoint}}
\put(680.0,869.0){\rule[-0.500pt]{30.353pt}{1.000pt}}
\put(806.0,869.0){\rule[-0.500pt]{1.000pt}{27.703pt}}
\put(680.0,984.0){\rule[-0.500pt]{30.353pt}{1.000pt}}
\put(680.0,869.0){\rule[-0.500pt]{1.000pt}{27.703pt}}
\put(1059,755){\usebox{\plotpoint}}
\put(1059.0,755.0){\rule[-0.500pt]{30.353pt}{1.000pt}}
\put(1185.0,755.0){\rule[-0.500pt]{1.000pt}{27.463pt}}
\put(1059.0,869.0){\rule[-0.500pt]{30.353pt}{1.000pt}}
\put(680,411){\raisebox{-.8pt}{\makebox(0,0){$\bigcirc$}}}
\put(806,411){\raisebox{-.8pt}{\makebox(0,0){$\bigcirc$}}}
\put(554,755){\raisebox{-.8pt}{\makebox(0,0){$\bigcirc$}}}
\put(554,869){\raisebox{-.8pt}{\makebox(0,0){$\bigcirc$}}}
\put(933,984){\raisebox{-.8pt}{\makebox(0,0){$\bigcirc$}}}
\put(1059,984){\raisebox{-.8pt}{\makebox(0,0){$\bigcirc$}}}
\put(1185,526){\raisebox{-.8pt}{\makebox(0,0){$\bigcirc$}}}
\put(1185,640){\raisebox{-.8pt}{\makebox(0,0){$\bigcirc$}}}
\put(933,411){\raisebox{-.8pt}{\makebox(0,0){$\bigcirc$}}}
\put(1059,411){\raisebox{-.8pt}{\makebox(0,0){$\bigcirc$}}}
\put(1059,526){\raisebox{-.8pt}{\makebox(0,0){$\bigcirc$}}}
\put(933,526){\raisebox{-.8pt}{\makebox(0,0){$\bigcirc$}}}
\put(933,411){\raisebox{-.8pt}{\makebox(0,0){$\bigcirc$}}}
\put(933,640){\raisebox{-.8pt}{\makebox(0,0){$\bigcirc$}}}
\put(933,755){\raisebox{-.8pt}{\makebox(0,0){$\bigcirc$}}}
\put(806,755){\raisebox{-.8pt}{\makebox(0,0){$\bigcirc$}}}
\put(806,640){\raisebox{-.8pt}{\makebox(0,0){$\bigcirc$}}}
\put(933,640){\raisebox{-.8pt}{\makebox(0,0){$\bigcirc$}}}
\put(554,526){\raisebox{-.8pt}{\makebox(0,0){$\bigcirc$}}}
\put(680,526){\raisebox{-.8pt}{\makebox(0,0){$\bigcirc$}}}
\put(680,640){\raisebox{-.8pt}{\makebox(0,0){$\bigcirc$}}}
\put(554,640){\raisebox{-.8pt}{\makebox(0,0){$\bigcirc$}}}
\put(554,526){\raisebox{-.8pt}{\makebox(0,0){$\bigcirc$}}}
\put(680,869){\raisebox{-.8pt}{\makebox(0,0){$\bigcirc$}}}
\put(806,869){\raisebox{-.8pt}{\makebox(0,0){$\bigcirc$}}}
\put(806,984){\raisebox{-.8pt}{\makebox(0,0){$\bigcirc$}}}
\put(680,984){\raisebox{-.8pt}{\makebox(0,0){$\bigcirc$}}}
\put(680,869){\raisebox{-.8pt}{\makebox(0,0){$\bigcirc$}}}
\put(1059,755){\raisebox{-.8pt}{\makebox(0,0){$\bigcirc$}}}
\put(1185,755){\raisebox{-.8pt}{\makebox(0,0){$\bigcirc$}}}
\put(1185,869){\raisebox{-.8pt}{\makebox(0,0){$\bigcirc$}}}
\put(1059,869){\raisebox{-.8pt}{\makebox(0,0){$\bigcirc$}}}
\put(1059,755){\raisebox{-.8pt}{\makebox(0,0){$\bigcirc$}}}
\put(1059.0,755.0){\rule[-0.500pt]{1.000pt}{27.463pt}}
\sbox{\plotpoint}{\rule[-0.250pt]{0.500pt}{0.500pt}}%
\put(680,640){\usebox{\plotpoint}}
\multiput(680,640)(12.453,0.000){11}{\usebox{\plotpoint}}
\put(806,640){\usebox{\plotpoint}}
\put(806,755){\usebox{\plotpoint}}
\multiput(806,755)(0.000,12.453){10}{\usebox{\plotpoint}}
\put(806,869){\usebox{\plotpoint}}
\put(933,640){\usebox{\plotpoint}}
\multiput(933,640)(0.000,-12.453){10}{\usebox{\plotpoint}}
\put(933,526){\usebox{\plotpoint}}
\put(933,755){\usebox{\plotpoint}}
\multiput(933,755)(12.453,0.000){11}{\usebox{\plotpoint}}
\put(1059,755){\usebox{\plotpoint}}
\put(680,526){\usebox{\plotpoint}}
\multiput(680,526)(0.000,-12.453){10}{\usebox{\plotpoint}}
\put(680,411){\usebox{\plotpoint}}
\put(806,411){\usebox{\plotpoint}}
\multiput(806,411)(12.453,0.000){11}{\usebox{\plotpoint}}
\put(933,411){\usebox{\plotpoint}}
\put(1059,526){\usebox{\plotpoint}}
\multiput(1059,526)(12.453,0.000){11}{\usebox{\plotpoint}}
\put(1185,526){\usebox{\plotpoint}}
\put(1185,640){\usebox{\plotpoint}}
\multiput(1185,640)(0.000,12.453){10}{\usebox{\plotpoint}}
\put(1185,755){\usebox{\plotpoint}}
\put(554,640){\usebox{\plotpoint}}
\multiput(554,640)(0.000,12.453){10}{\usebox{\plotpoint}}
\put(554,755){\usebox{\plotpoint}}
\put(554,869){\usebox{\plotpoint}}
\multiput(554,869)(12.453,0.000){11}{\usebox{\plotpoint}}
\put(680,869){\usebox{\plotpoint}}
\put(806,984){\usebox{\plotpoint}}
\multiput(806,984)(12.453,0.000){11}{\usebox{\plotpoint}}
\put(933,984){\usebox{\plotpoint}}
\put(1059,869){\usebox{\plotpoint}}
\multiput(1059,869)(0.000,12.453){10}{\usebox{\plotpoint}}
\put(1059,984){\usebox{\plotpoint}}
\put(1059,411){\usebox{\plotpoint}}
\multiput(1059,411)(0.000,-12.453){6}{\usebox{\plotpoint}}
\put(1059,337){\usebox{\plotpoint}}
\put(554,526){\usebox{\plotpoint}}
\multiput(554,526)(-12.453,0.000){7}{\usebox{\plotpoint}}
\put(472,526){\usebox{\plotpoint}}
\put(680,984){\usebox{\plotpoint}}
\multiput(680,984)(0.000,12.453){6}{\usebox{\plotpoint}}
\put(680,1058){\usebox{\plotpoint}}
\put(1185,869){\usebox{\plotpoint}}
\multiput(1185,869)(12.453,0.000){7}{\usebox{\plotpoint}}
\put(1267,869){\usebox{\plotpoint}}
\sbox{\plotpoint}{\rule[-0.500pt]{1.000pt}{1.000pt}}%
\put(680,411){\usebox{\plotpoint}}
\put(680.0,337.0){\rule[-0.500pt]{1.000pt}{17.827pt}}
\put(806,411){\usebox{\plotpoint}}
\put(806.0,337.0){\rule[-0.500pt]{1.000pt}{17.827pt}}
\put(1185,526){\usebox{\plotpoint}}
\put(1185.0,526.0){\rule[-0.500pt]{19.754pt}{1.000pt}}
\put(1185,640){\usebox{\plotpoint}}
\put(1185.0,640.0){\rule[-0.500pt]{19.754pt}{1.000pt}}
\put(1059,984){\usebox{\plotpoint}}
\put(1059.0,984.0){\rule[-0.500pt]{1.000pt}{17.827pt}}
\put(933,984){\usebox{\plotpoint}}
\put(933.0,984.0){\rule[-0.500pt]{1.000pt}{17.827pt}}
\put(554,755){\usebox{\plotpoint}}
\put(472.0,755.0){\rule[-0.500pt]{19.754pt}{1.000pt}}
\put(554,869){\usebox{\plotpoint}}
\put(472.0,869.0){\rule[-0.500pt]{19.754pt}{1.000pt}}
\sbox{\plotpoint}{\rule[-0.175pt]{0.350pt}{0.350pt}}%
\put(302,182){\usebox{\plotpoint}}
\put(302.0,182.0){\rule[-0.175pt]{0.350pt}{248.368pt}}
\put(302.0,1213.0){\rule[-0.175pt]{273.421pt}{0.350pt}}
\put(1437.0,182.0){\rule[-0.175pt]{0.350pt}{248.368pt}}
\put(302.0,182.0){\rule[-0.175pt]{273.421pt}{0.350pt}}
\end{picture}

\newpage
\setlength{\unitlength}{0.240900pt}
\ifx\plotpoint\undefined\newsavebox{\plotpoint}\fi
\sbox{\plotpoint}{\rule[-0.500pt]{1.000pt}{1.000pt}}%
\begin{picture}(900,1300)(90,100)
\font\gnuplot=cmr10 at 10pt
\gnuplot
\sbox{\plotpoint}{\rule[-0.500pt]{1.000pt}{1.000pt}}%
\put(1058,940){\makebox(0,0)[l]{$J_1$}}
\put(1112,770){\makebox(0,0)[l]{$\longleftarrow$}}
\put(1200,770){\makebox(0,0)[1]{$J_2$}}
\put(880,726){\makebox(0,0)[1]{$J_3$}}
\put(810,173){\makebox(0,0)[l]{{\large Fig. 2}}}
\put(760,850){\makebox(0,0)[l]{$J_4$}}
\put(869,561){\makebox(0,0)[l]{$J_5$}}
\put(666,582){\makebox(0,0)[l]{$J_6$}}
\put(572,488){\usebox{\plotpoint}}
\put(572.0,488.0){\rule[-0.500pt]{47.698pt}{1.000pt}}
\put(770.0,488.0){\rule[-0.500pt]{1.000pt}{50.348pt}}
\put(572.0,697.0){\rule[-0.500pt]{47.698pt}{1.000pt}}
\put(572.0,488.0){\rule[-0.500pt]{1.000pt}{50.348pt}}
\put(969,697){\usebox{\plotpoint}}
\put(969.0,697.0){\rule[-0.500pt]{47.698pt}{1.000pt}}
\put(1167.0,697.0){\rule[-0.500pt]{1.000pt}{50.589pt}}
\put(969.0,907.0){\rule[-0.500pt]{47.698pt}{1.000pt}}
\put(572,488){\raisebox{-.8pt}{\makebox(0,0){$\bigcirc$}}}
\put(770,488){\raisebox{-.8pt}{\makebox(0,0){$\bigcirc$}}}
\put(770,697){\raisebox{-.8pt}{\makebox(0,0){$\bigcirc$}}}
\put(572,697){\raisebox{-.8pt}{\makebox(0,0){$\bigcirc$}}}
\put(572,488){\raisebox{-.8pt}{\makebox(0,0){$\bigcirc$}}}
\put(969,697){\raisebox{-.8pt}{\makebox(0,0){$\bigcirc$}}}
\put(1167,697){\raisebox{-.8pt}{\makebox(0,0){$\bigcirc$}}}
\put(1167,907){\raisebox{-.8pt}{\makebox(0,0){$\bigcirc$}}}
\put(969,907){\raisebox{-.8pt}{\makebox(0,0){$\bigcirc$}}}
\put(969,697){\raisebox{-.8pt}{\makebox(0,0){$\bigcirc$}}}
\put(969.0,697.0){\rule[-0.500pt]{1.000pt}{50.589pt}}
\sbox{\plotpoint}{\rule[-0.250pt]{0.500pt}{0.500pt}}%
\put(572,697){\usebox{\plotpoint}}
\multiput(572,697)(7.593,9.871){2}{\usebox{\plotpoint}}
\put(587.72,716.29){\usebox{\plotpoint}}
\put(596.10,725.51){\usebox{\plotpoint}}
\put(604.48,734.72){\usebox{\plotpoint}}
\multiput(612,743)(8.806,8.806){2}{\usebox{\plotpoint}}
\put(630.94,761.05){\usebox{\plotpoint}}
\put(640.61,768.89){\usebox{\plotpoint}}
\put(650.34,776.67){\usebox{\plotpoint}}
\put(660.15,784.34){\usebox{\plotpoint}}
\put(670.32,791.52){\usebox{\plotpoint}}
\put(680.97,797.98){\usebox{\plotpoint}}
\multiput(681,798)(11.139,5.569){0}{\usebox{\plotpoint}}
\put(692.10,803.55){\usebox{\plotpoint}}
\put(703.33,808.93){\usebox{\plotpoint}}
\put(714.89,813.56){\usebox{\plotpoint}}
\put(726.76,817.15){\usebox{\plotpoint}}
\put(738.79,820.34){\usebox{\plotpoint}}
\multiput(741,821)(12.391,1.239){0}{\usebox{\plotpoint}}
\put(751.09,822.01){\usebox{\plotpoint}}
\put(763.50,823.00){\usebox{\plotpoint}}
\put(775.95,823.00){\usebox{\plotpoint}}
\put(788.36,822.16){\usebox{\plotpoint}}
\multiput(790,822)(12.391,-1.239){0}{\usebox{\plotpoint}}
\put(800.72,820.78){\usebox{\plotpoint}}
\put(812.71,817.46){\usebox{\plotpoint}}
\put(824.66,814.13){\usebox{\plotpoint}}
\put(836.23,809.51){\usebox{\plotpoint}}
\put(847.50,804.25){\usebox{\plotpoint}}
\put(858.64,798.68){\usebox{\plotpoint}}
\put(869.38,792.37){\usebox{\plotpoint}}
\multiput(870,792)(9.830,-7.646){0}{\usebox{\plotpoint}}
\put(879.27,784.81){\usebox{\plotpoint}}
\put(889.45,777.64){\usebox{\plotpoint}}
\multiput(899,770)(9.724,-7.780){2}{\usebox{\plotpoint}}
\put(918.16,753.76){\usebox{\plotpoint}}
\put(927.00,745.00){\usebox{\plotpoint}}
\put(935.48,735.87){\usebox{\plotpoint}}
\put(943.85,726.66){\usebox{\plotpoint}}
\put(952.23,717.44){\usebox{\plotpoint}}
\multiput(959,710)(7.593,-9.871){2}{\usebox{\plotpoint}}
\put(969,697){\usebox{\plotpoint}}
\put(572,488){\usebox{\plotpoint}}
\multiput(572,488)(11.020,5.801){37}{\usebox{\plotpoint}}
\put(969,697){\usebox{\plotpoint}}
\put(770,697){\usebox{\plotpoint}}
\multiput(770,697)(12.453,0.000){16}{\usebox{\plotpoint}}
\put(969,697){\usebox{\plotpoint}}
\put(770,488){\usebox{\plotpoint}}
\multiput(770,488)(8.587,9.019){24}{\usebox{\plotpoint}}
\put(969,697){\usebox{\plotpoint}}
\sbox{\plotpoint}{\rule[-0.175pt]{0.350pt}{0.350pt}}%
\put(1167,697){\usebox{\plotpoint}}
\multiput(1166.02,697.00)(-0.500,0.531){393}{\rule{0.120pt}{0.459pt}}
\multiput(1166.27,697.00)(-198.000,209.048){2}{\rule{0.350pt}{0.229pt}}
\put(969,697){\usebox{\plotpoint}}
\multiput(969.48,697.00)(0.500,0.531){393}{\rule{0.120pt}{0.459pt}}
\multiput(968.27,697.00)(198.000,209.048){2}{\rule{0.350pt}{0.229pt}}
\put(374,278){\usebox{\plotpoint}}
\put(374.0,278.0){\rule[-0.175pt]{238.732pt}{0.350pt}}
\put(1365.0,278.0){\rule[-0.175pt]{0.350pt}{202.115pt}}
\put(374.0,1117.0){\rule[-0.175pt]{238.732pt}{0.350pt}}
\put(374.0,278.0){\rule[-0.175pt]{0.350pt}{202.115pt}}
\end{picture}

\newpage
\setlength{\unitlength}{0.240900pt}
\ifx\plotpoint\undefined\newsavebox{\plotpoint}\fi
\sbox{\plotpoint}{\rule[-0.500pt]{1.000pt}{1.000pt}}%
\begin{picture}(1000,1300)(90,0)
\font\gnuplot=cmr10 at 10pt
\gnuplot
\sbox{\plotpoint}{\rule[-0.500pt]{1.000pt}{1.000pt}}%
\put(407,739){\makebox(0,0)[l]{i}}
\put(638,739){\makebox(0,0)[l]{j}}
\put(638,440){\makebox(0,0)[l]{k}}
\put(407,440){\makebox(0,0)[l]{l}}
\put(1101,739){\makebox(0,0)[l]{i}}
\put(1332,739){\makebox(0,0)[l]{j}}
\put(1332,440){\makebox(0,0)[l]{k}}
\put(1101,440){\makebox(0,0)[l]{l}}
\put(830,593){\makebox(0,0)[l]{$+$}}
\put(754,173){\makebox(0,0)[l]{{\large Fig. 3}}}
\put(523,698){\makebox(0,0)[l]{$>$}}
\put(523,488){\makebox(0,0)[l]{$<$}}
\put(1087,593){\makebox(0,0)[l]{$\bigvee$}}
\put(1317,593){\makebox(0,0)[l]{$\bigwedge$}}
\put(407,488){\usebox{\plotpoint}}
\put(407.0,488.0){\rule[-0.500pt]{55.648pt}{1.000pt}}
\put(407,698){\usebox{\plotpoint}}
\put(407.0,698.0){\rule[-0.500pt]{55.648pt}{1.000pt}}
\put(1101,488){\usebox{\plotpoint}}
\put(1101.0,488.0){\rule[-0.500pt]{1.000pt}{50.589pt}}
\put(1332,488){\usebox{\plotpoint}}
\put(407,488){\raisebox{-.8pt}{\makebox(0,0){$\bigcirc$}}}
\put(638,488){\raisebox{-.8pt}{\makebox(0,0){$\bigcirc$}}}
\put(407,698){\raisebox{-.8pt}{\makebox(0,0){$\bigcirc$}}}
\put(638,698){\raisebox{-.8pt}{\makebox(0,0){$\bigcirc$}}}
\put(1101,488){\raisebox{-.8pt}{\makebox(0,0){$\bigcirc$}}}
\put(1101,698){\raisebox{-.8pt}{\makebox(0,0){$\bigcirc$}}}
\put(1332,488){\raisebox{-.8pt}{\makebox(0,0){$\bigcirc$}}}
\put(1332,698){\raisebox{-.8pt}{\makebox(0,0){$\bigcirc$}}}
\put(1332.0,488.0){\rule[-0.500pt]{1.000pt}{50.589pt}}
\sbox{\plotpoint}{\rule[-0.175pt]{0.350pt}{0.350pt}}%
\put(176,278){\usebox{\plotpoint}}
\put(176.0,278.0){\rule[-0.175pt]{334.128pt}{0.350pt}}
\put(1563.0,278.0){\rule[-0.175pt]{0.350pt}{151.526pt}}
\put(176.0,907.0){\rule[-0.175pt]{334.128pt}{0.350pt}}
\put(176.0,278.0){\rule[-0.175pt]{0.350pt}{151.526pt}}
\end{picture}
\newpage

\setlength{\unitlength}{0.240900pt}
\ifx\plotpoint\undefined\newsavebox{\plotpoint}\fi
\sbox{\plotpoint}{\rule[-0.500pt]{1.000pt}{1.000pt}}%
\begin{picture}(900,1300)(90,0)
\font\gnuplot=cmr10 at 10pt
\gnuplot
\sbox{\plotpoint}{\rule[-0.500pt]{1.000pt}{1.000pt}}%
\put(716,967){\makebox(0,0)[l]{i}}
\put(716,787){\makebox(0,0)[l]{j}}
\put(969,833){\makebox(0,0)[l]{k}}
\put(1167,833){\makebox(0,0)[l]{l}}
\put(560,565){\makebox(0,0)[l]{m}}
\put(760,565){\makebox(0,0)[l]{n}}
\put(1010,608){\makebox(0,0)[l]{o}}
\put(1010,428){\makebox(0,0)[l]{p}}
\put(810,158){\makebox(0,0)[l]{{\large Fig. 4}}}
\put(572,608){\usebox{\plotpoint}}
\put(572.0,608.0){\rule[-0.500pt]{95.637pt}{1.000pt}}
\put(969.0,608.0){\rule[-0.500pt]{1.000pt}{43.121pt}}
\put(770.0,787.0){\rule[-0.500pt]{47.939pt}{1.000pt}}
\put(770.0,608.0){\rule[-0.500pt]{1.000pt}{43.121pt}}
\put(969,428){\usebox{\plotpoint}}
\put(969.0,428.0){\rule[-0.500pt]{1.000pt}{43.362pt}}
\put(1167,787){\usebox{\plotpoint}}
\put(969.0,787.0){\rule[-0.500pt]{47.698pt}{1.000pt}}
\put(770,967){\usebox{\plotpoint}}
\put(572,608){\raisebox{-.8pt}{\makebox(0,0){$\bigcirc$}}}
\put(770,608){\raisebox{-.8pt}{\makebox(0,0){$\bigcirc$}}}
\put(969,608){\raisebox{-.8pt}{\makebox(0,0){$\bigcirc$}}}
\put(969,787){\raisebox{-.8pt}{\makebox(0,0){$\bigcirc$}}}
\put(770,787){\raisebox{-.8pt}{\makebox(0,0){$\bigcirc$}}}
\put(770,608){\raisebox{-.8pt}{\makebox(0,0){$\bigcirc$}}}
\put(969,428){\raisebox{-.8pt}{\makebox(0,0){$\bigcirc$}}}
\put(969,608){\raisebox{-.8pt}{\makebox(0,0){$\bigcirc$}}}
\put(1167,787){\raisebox{-.8pt}{\makebox(0,0){$\bigcirc$}}}
\put(969,787){\raisebox{-.8pt}{\makebox(0,0){$\bigcirc$}}}
\put(770,967){\raisebox{-.8pt}{\makebox(0,0){$\bigcirc$}}}
\put(770,787){\raisebox{-.8pt}{\makebox(0,0){$\bigcirc$}}}
\put(770.0,787.0){\rule[-0.500pt]{1.000pt}{43.362pt}}
\sbox{\plotpoint}{\rule[-0.175pt]{0.350pt}{0.350pt}}%
\put(374,248){\usebox{\plotpoint}}
\put(374.0,248.0){\rule[-0.175pt]{238.732pt}{0.350pt}}
\put(1365.0,248.0){\rule[-0.175pt]{0.350pt}{216.569pt}}
\put(374.0,1147.0){\rule[-0.175pt]{238.732pt}{0.350pt}}
\put(374.0,248.0){\rule[-0.175pt]{0.350pt}{216.569pt}}
\end{picture}

\end{document}